\documentclass{DISproc}

\pdfoutput=1

\usepackage{amssymb}
\usepackage{amsmath}
\usepackage{graphicx}
\usepackage{slashed}

\newcommand{\slv}{\raise.15ex\hbox{$/$}\kern-.53em\hbox{$v$}}
\newcommand{\slF}{\raise.15ex\hbox{$/$}\kern-.53em\hbox{$F$}}
\newcommand{\slL}{\raise.15ex\hbox{$/$}\kern-.53em\hbox{$L$}}
\newcommand{\slP}{\raise.15ex\hbox{$/$}\kern-.53em\hbox{$P$}}
\newcommand{\slp}{\raise.15ex\hbox{$/$}\kern-.53em\hbox{$p$}}
\newcommand{\slq}{\raise.15ex\hbox{$/$}\kern-.53em\hbox{$q$}}
\newcommand{\slR}{\raise.15ex\hbox{$/$}\kern-.53em\hbox{$R$}}
\newcommand{\slQ}{\raise.15ex\hbox{$/$}\kern-.53em\hbox{$Q$}}
\newcommand{\slK}{\raise.15ex\hbox{$/$}\kern-.53em\hbox{$K$}}
\newcommand{\slk}{\raise.15ex\hbox{$/$}\kern-.53em\hbox{$k$}}
\newcommand{\slD}{\raise.15ex\hbox{$/$}\kern-.53em\hbox{$D$}}
\newcommand{\slC}{\raise.15ex\hbox{$/$}\kern-.53em\hbox{$C$}}
\newcommand{\slA}{\raise.15ex\hbox{$/$}\kern-.53em\hbox{$A$}}
\newcommand{\slSigma}{\raise.15ex\hbox{$/$}\kern-.53em\hbox{$\Sigma$}}
\newcommand{\slpartial}{\raise.15ex\hbox{$/$}\kern-.53em\hbox{$\partial$}}
\newcommand{\slcalP}{\raise.15ex\hbox{$/$}\kern-.63em\hbox{$\cal P$}}

\def\bs{\boldsymbol}

\def\bx{\bar x}

\def\pk{p\cdot k}
\def\hk{\bar{p} \cdot k}
\def\pv{p\cdot v}
\def\hv{\bar {p}\cdot v}
\def\q{{\boldsymbol q}}

\def\k{{\boldsymbol k}}
\def\kh{{\boldsymbol \kappa}}
\def\khb{\bar{{\boldsymbol \kappa}}}
\def\Qh{{\boldsymbol \nu}}
\def\Qhb{\bar{{\boldsymbol \nu}}}

\def\x{{\boldsymbol x}}

\def\D{{\boldsymbol D}_\perp}

\def\Oh{\Omega_{\bar{p}}}

\def\D10{\Delta^+_{10}}

\def\x+{x^+}

\newcommand{\beq}{\begin{eqnarray}}
\newcommand{\eeq}{\end{eqnarray}}
\newcommand{\be}{\begin{equation}}
\newcommand{\ee}{\end{equation}}

\begin{document}

\title{Medium-induced soft gluon radiation in DIS}

\author{{\slshape N\'estor Armesto$^1$, Hao Ma$^1$, Mauricio Mart\'inez$^1$, Yacine Mehtar-Tani$^2$, Carlos A. Salgado$^1$}\\[1ex]
$^1$Departamento de F\'isica de Part\'iculas and IGFAE,
Universidade de Santiago de Compostela \\
E-15782 Santiago de Compostela, 
Galicia-Spain\\
$^2$Institut de Physique Th\'eorique, CEA Saclay, 
F-91191 Gif-sur-Yvette, France}

\contribID{0}

\doi

\maketitle

\begin{abstract}
We study color coherence effects on the medium-induced soft gluon radiation off an asymptotic quark hit by a virtual photon traversing a hot and dense QCD medium. The transverse momentum spectrum of the emitted gluon is computed at 1st order in the opacity expansion. The interference effects between the initial and final state radiation modify the soft gluon spectrum when a finite angle between the incoming and outgoing quarks is considered, presenting a soft divergence. We comment on possible implications on observables in eA collisions.
\end{abstract}

\section{Introduction}
Color coherence effects in vacuum are found by TASSO and OPAL experiments \cite{Braunschweig:1990yd, Abbiendi:2002mj}, by the depletion of particle spectrum for small energies. Medium-induced soft gluon radiation off a single quark in the final state was also considered long ago \cite{Baier:1996kr, Zakharov:1997uu, Wiedemann:2000za, Gyulassy:2000er}. Our goal is to check the color coherence effects between initial and final states in the presence of a medium which could be of applicability in DIS on nucleus. It is a setup complementary to the antenna in the $s$-channel \cite{MehtarTani:2010ma, CasalderreySolana:2011rz, Armesto:2011ir, MehtarTani:2011tz}. In high energy DIS the virtual photon scatters on one quark inside the nucleus to change its transverse momentum, and then such quark can rescatter on the fields generated by the other components of the nucleus. The process we consider is just a first step in studying eA collisions \footnote{It still needs to be further improved. If the exchanging particle is a highly virtual gluon, then one can consider the process as pA collisions.}, which will be studied in the future LHeC and EIC colliders. When the scattering angle between the incoming and outgoing quarks becomes 0, our calculation matches the one for gluon production in the totally coherent limit in the CGC framework \cite{Kovchegov:1998bi}.

\section{The antenna spectrum in $t$-channel in vacuum}

\begin{figure*}
\begin{center}
\includegraphics[width=0.8\textwidth]{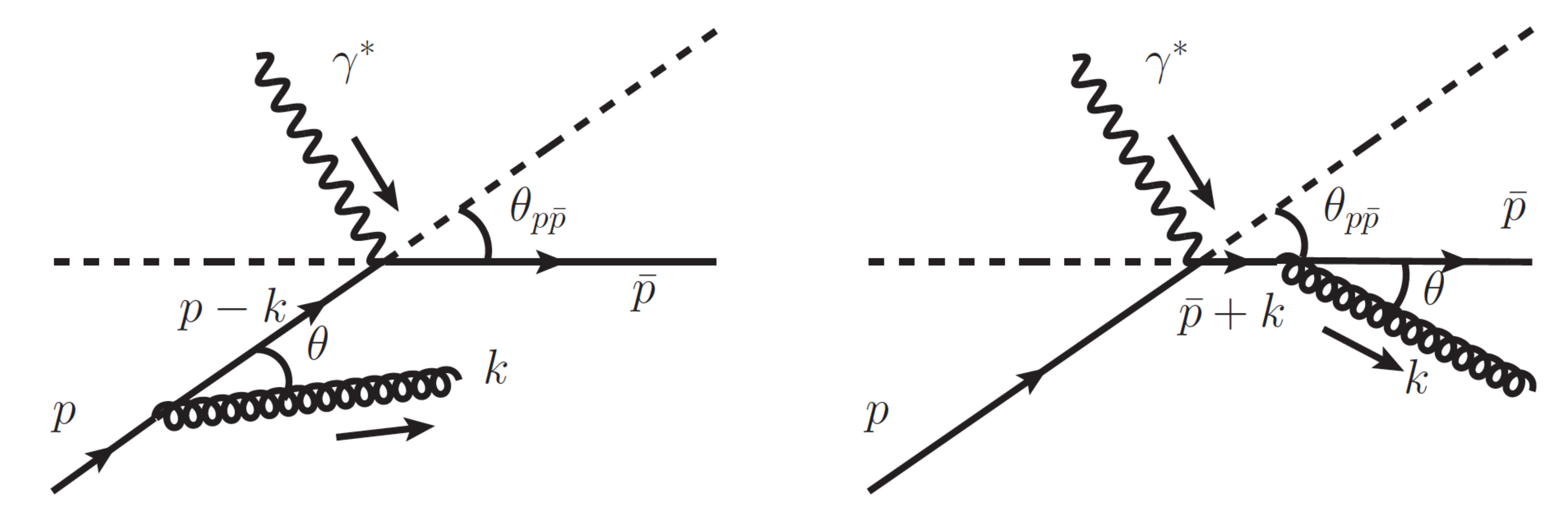}
\end{center}
\caption{Antenna radiation in $t$-channel in vacuum}
\label{fig:vacuum}
\end{figure*}

Antenna radiation in $t$-channel in vacuum is shown in Fig.\ref{fig:vacuum}. We work in the infinite momentum frame. Small scattering angle between incoming and outgoing quarks ($p$ and $\bar p$, respectively) is assumed, i.e. $\theta_{p {\bar p}} \ll 1$. The eikonal approximation is employed, i.e. $p^+ \sim {\bar p}^+ \gg k^+ \gg |\bf k|$, where the relation between the forward light-cone momentum and the energy of the emitted soft gluon reads $k^+ = \sqrt{2} \, \omega$. The light-cone gauge $n \cdot A = A^+ = 0$ is chosen, with the axial vector $n = (0, 1, {\bf 0})$. A virtual photon $\gamma^*$ with transverse size $|\Delta {\bf x}| \sim 1 / |{\bar {\bf p}} - {\bf p}|$ is absorbed by an incoming quark. The soft gluon spectrum off the incoming quark in vacuum in which the azimuthal angle is integrated with respect to the direction of quark (analogously for the outgoing quark) reads
\begin{equation}
d N^{\rm vac}_{\rm in} = \frac{\alpha_s \, C_F}{\pi} \, \frac{d \omega}{\omega} \, \frac{\sin\theta \, d \theta}{1 - \cos\theta} \, \Theta (\cos\theta - \cos \theta_{p \bar{p}}),
\end{equation}
where $\alpha_s$ is the strong coupling constant, $C_F$ is the Casimir factor of the fundamental representation, $\theta \approx |\k| / \omega$ is the gluon emission angle, and the heaviside step function gives the angular constraint $\theta < \theta_{p {\bar p}}$, i.e. the soft gluon emission is constrained to be inside the cone set by the scattering angle between the incoming and outgoing quarks \cite{Basics of Perturbative QCD}. As long as $|{\boldsymbol \lambda}| = 1 / |\k|$, the transverse wave length of the emitted gluon, is shorter than $|{\boldsymbol r}| \sim t_{\rm form} \, \theta_{p {\bar p}}$ (the change of position of the quark in the transverse plane due to the photon scattering when the gluon is formed \footnote{The formation time is defined as $t_{\rm form} \sim \omega / \k^2$.}), the gluon resolves the color structure and the bremsstrahlung is therefore off either the incoming quark or the outgoing quark. In terms of angles, one gets immediately $\theta < \theta_{p {\bar p}}$ from the simple analysis above.

\section{The medium-induced antenna spectrum in $t$-channel}

\begin{figure*}
\begin{center}
\includegraphics[width=0.8\textwidth]{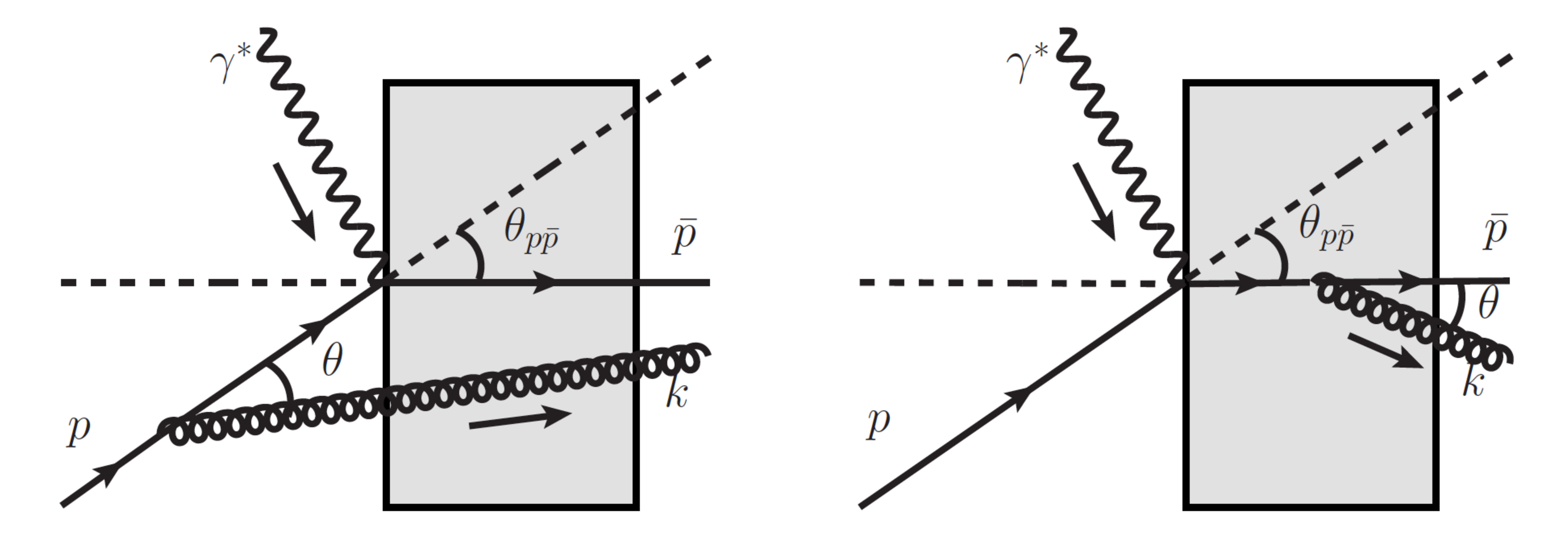}
\end{center}
\caption{Medium-induced antenna radiation in $t$-channel}
\label{fig:medium}
\end{figure*}

Medium-induced antenna radiation in $t$-channel is shown in Fig.\ref{fig:medium}. We work in the same setup as the one in vacuum. Medium is assumed to act right after the photon scattering. Note that there is no need to go into the details of how the medium is modeled in our study. We consider dilute medium scenario, i.e. one gluon exchange. The opacity is defined as $L / \lambda$, and when we say ``opacity expansion'', we mean the expansion in number of scatterings. Hadronization is considered to occur outside the medium, therefore the process we consider is a pure perturbative one. The medium-induced antenna spectrum in $t$-channel at 1st order in opacity expansion reads:
\begin{equation}\label{eq:Total_spectrum}
\begin{split}
\omega\frac{dN^\text{med}}{d^3\vec{k}} = & \, \frac{\alpha_s C_F \,\hat q}{\pi} \int \frac{d^2 \q}{(2\pi)^2}{\cal V}^2(\q)~\int_0^{+ \infty} dx^+\\
& \, \Biggl[\frac{\Qh^2}{x^2\,(\pv)^2}-\frac{\kh^2}{x^2\,(\pk)^2}\\
& \, + \frac{2}{\bx^2}\Biggl(\frac{\Qhb^2}{(\hv)^2}-\frac{\khb\cdot\Qhb}{(\hv)(\hk)}\Biggr)(1-\cos\bigl[\Oh\x+\bigr])\\
& \, + \frac{2}{x\,\bar{x}}\,\Biggl(\frac{\kh\cdot\khb}{(\hk)\,(\pk)}\,-\frac{\Qh\cdot\khb}{(\hk)\,(\pv)}\\
& \, + \Bigl(\frac{\Qh\cdot\khb}{(\hk)\,(\pv)}-\,\frac{\Qh\cdot\Qhb}{(\hv)\,(\pv)}\Bigr)\,\bigl(1-\cos\bigl[\Oh\x+\bigr])\Biggr)\Biggr],
\end{split}
\end{equation}
where ${\hat q}=\alpha_s C_A n_0 m_D^2$ denotes the medium transport coefficient and $\Omega_{\bar p} = ({\bf k} - {\bf q})^2 / (2 \, k^+)$ is the inverse of the gluon formation length. The second line in Eq.(\ref{eq:Total_spectrum}) is the contribution from the gluon radiation off the initial quark. The third line in Eq.(\ref{eq:Total_spectrum}) comes from the medium-induced gluon radiation off the outgoing quark, i.e. the independent spectrum \cite{Armesto:2011ir}. The last two lines of Eq.(\ref{eq:Total_spectrum}) is the interference between the incoming and outgoing quarks. Note that when the scattering angle $\theta_{p {\bar p}} = 0$, the spectrum becomes \footnote{The case of $\theta_{p {\bar p}} = 0$ was studied in the multiple soft scattering approach in \cite{Kovchegov:1998bi}.}
\begin{equation}\label{eq:0_scattering_angle}
\omega\frac{dN^{\rm med}}{d^3\vec{k}} = \frac{4 \, \alpha_s C_F \, {\hat q} \, L^+}{\pi} \int \frac{d^2 {\bf q}}{(2\pi)^2}{\cal V}^2({\bf q}) \, {\boldsymbol L}^2,
\end{equation}
where ${\bs L}= (\kh - \q) / (\kh - \q)^2 - \kh / \kh^2$ is the transverse component of the Lipatov vertex in the light-cone gauge. The notations are $\kh = \k - x \, {\boldsymbol p}$, $x = k^+ / p^+$ and $\khb = \k - {\bar x} \, {\bar {\boldsymbol p}}$, ${\bar x} = k^+ / {\bar p}^+$. The structure of the transverse component of the gauge invariant Lipatov vertex indicates that Eq.(\ref{eq:0_scattering_angle}) is the genuine medium-induced gluon radiation off an on-shell quark which comes from the $- \infty$ and goes to the $+\infty$.

\section{Soft limit}
In the soft gluon emission limit ($\omega \rightarrow 0$), the antenna spectrum in $t$-channel in medium, adding the medium-induced and the vacuum contributions, reads
\begin{equation}\label{eq:Soft_limit}
\omega \frac{d N^{\rm vac}}{d^3 \vec{k}} + \omega \frac{d N^{\rm med}}{d^3 \vec{k}} = \frac{4 \, \alpha_s \, C_F}{( 2 \, \pi )^2} \left[ (1 - \Delta) \left( \frac{1}{{\boldsymbol \kappa}^2} - \frac{{\boldsymbol \kappa} \cdot \bar{{\boldsymbol \kappa}}}{{\boldsymbol \kappa}^2 \, \bar{{\boldsymbol \kappa}}^2} \right) + \frac{1}{\bar{{\boldsymbol \kappa}}^2} - (1 - \Delta) \, \frac{{\boldsymbol \kappa} \cdot \bar{{\boldsymbol \kappa}}}{{\boldsymbol \kappa}^2 \, \bar{{\boldsymbol \kappa}}^2} \right],
\end{equation}
where $\Delta = {\hat q} \, L^+ / m_D^2$ at 1st order in opacity. The first term in the brackets shows the angular constraint for a reduced number of soft gluon emission off the incoming quark if one performs the azimuthal angle integration for the emitted soft gluon. The reason for the reduction of the soft gluon multiplicity inside the cone is that, the emitted soft gluon off the incoming quark will suffer rescattering when it goes through the medium. It therefore slows down the evolution of the gluon density. The rest in the brackets is the soft gluon emission off the outgoing quark in medium. When the medium is switched off, i.e. $\Delta \rightarrow 0$, one naturally gets the vacuum contribution. In the opaque medium limit, i.e. $\Delta \rightarrow 1$, one has
\begin{equation}\label{eq:opaque_medium}
\omega \frac{d N^{\rm vac}}{d^3 \vec{k}} + \omega \frac{d N^{\rm med}}{d^3 \vec{k}} = \frac{4 \, \alpha_s \, C_F}{( 2 \, \pi )^2} \, \frac{1}{\bar{{\boldsymbol \kappa}}^2}.
\end{equation}
After comparing Eq.(\ref{eq:opaque_medium}) with Eq.(\ref{eq:Soft_limit}), one can see that the soft part of the medium-induced gluon energy spectrum off the incoming quark is suppressed, i.e. the gluon density is saturated, and the bremsstralung contribution between vacuum and medium-induced parts get canceled with each other. In Eq.(\ref{eq:opaque_medium}), one gets complete color decoherence for the outgoing quark, i.e. in the opaque medium the outgoing quark loses the color coherence with the incoming quark, and then the soft gluon radiation off the outgoing quark is like the soft gluon radiation off a single quark in vacuum. A similar property was discovered in the antenna spectrum in $s$-channel \cite{Armesto:2011ir, MehtarTani:2011tz}.

Generalizing the results to the multiple soft scattering limit is in progress.

{\raggedright
\begin{footnotesize}

\end{footnotesize}
}

\end{document}